\documentclass[preprint]{elsarticle}
\journal{Nuclear Instruments and Methods in Physics Research A}

\usepackage{amssymb}

\usepackage{float}

\usepackage{lineno}

\usepackage{myjournal}

\usepackage[margin=1in]{geometry}

\begin{document}

\begin{frontmatter}



\title{A Novel \emph{in situ} Trigger Combination Method}

\author[mcgill]{Adrian~Buzatu\corref{cor1}\fnref{fn1}}
\ead{adrian.buzatu@mail.mcgill.ca}
\author[mcgill]{Andreas~Warburton}
\ead{andreas.warburton@mcgill.ca}
\author[baylor]{Nils~Krumnack\fnref{fn3}}
\ead{nils@fnal.gov}
\author[lbnl]{Wei-Ming~Yao}
\ead{wmyao@lbl.gov}

\cortext[cor1]{Corresponding author}
\fntext[fn1]{Present address: University of Glasgow, Glasgow, UK, G12 8QQ; Tel: +44(0)141 330 2599; Fax: +44(0)141 330 5881}
\fntext[fn3]{Present address: Iowa State University, Ames, Iowa, USA, 50011}

\address[mcgill]{McGill University, Montr\'eal, Qu\'ebec, Canada, H3A\,2T8}
\address[baylor]{Baylor University, Waco, Texas, USA, 76798}
\address[lbnl]{Ernest Orlando Lawrence Berkeley National Laboratory, Berkeley, California, USA, 94720}

\begin{abstract}

\ \\Searches for rare physics processes using particle detectors in high-luminosity colliding hadronic beam environments require the use of multi-level trigger systems to reject colossal background rates in real time.  In analyses like the search for the Higgs boson, there is a need to maximize the signal acceptance by combining multiple different trigger chains when forming the offline data sample. In such statistically limited searches, datasets are often amassed over periods of several years, during which the trigger characteristics evolve and system performance can vary significantly. Reliable production cross-section measurements and upper limits must take into account a detailed understanding of the effective trigger inefficiency for every selected candidate event. We present as an example the complex situation of three trigger chains, based on missing energy and jet energy, that were combined in the context of the search for the
Higgs ($H$) boson produced in association with a $W$ boson at the Collider Detector at Fermilab (CDF). We briefly review the existing techniques for combining triggers, namely the inclusion, division, and exclusion methods.  We introduce and describe a novel fourth \emph{in situ} method whereby, for each candidate event, only the trigger chain with the highest \emph{a priori} probability of selecting the event is considered.  We compare the inclusion and novel \emph{in situ} methods for signal event yields in the CDF $WH$ search. This new combination method, by virtue of its scalability to large numbers of differing trigger chains and insensitivity to correlations between triggers, will benefit future long-running collider experiments, including those currently operating on the Large Hadron Collider.
\end{abstract}

\end{frontmatter}

\tableofcontents


\section{Introduction}
\label{section:Introductions}

\ \\Today's particle physics detectors at colliders use multi-level trigger systems to select online only a small fraction of the total number of events occurring each second.  A typical offline analysis compares the total event yield prediction for background and signal processes with the data event yield. In Monte-Carlo-simulated (MC) background or signal events, one needs to account for the fact that not all existing data events are selected by the trigger. In order to maximize the statistical power of the available MC events, in some experiments the trigger is not simulated in MC events with a binary 'accept' or 'reject' choice, as for data events, but rather all MC events are kept and assigned a probability that the trigger selects the particular event based on the MC event kinematic information and the trigger parametrization computed in data events. The trigger acceptance rate is often artificially reduced in order to cope with bandwidth problems. Such a prescaling is modelled in MC events by including the inverse of the prescale factor in the probability assigned to the event. For statistically limited analyses, there is often a need to maximize the signal event yield by combining several different kinds of triggers. Combining several triggers, however, presents additional challenges when not all triggers were entirely active or functioning properly while forming the desired dataset, or when each trigger selection may be required only for a certain set of offline event selection criteria. 

\ \\We begin this paper by presenting in Section~\ref{section:Definitions} the trigger definitions we employ. We continue in Section~\ref{section:Triggers} by describing the illustrative example of three triggers, based on missing transverse energy and jets, available in the full dataset collected over several years of Tevatron $p\bar{p}$ collisions by the Collider Detector at Fermilab (CDF). During this period the trigger configurations changed several times, new triggers were introduced, and some triggers did not function properly for a fraction of the total integrated luminosity. These three triggers, based on signatures of a charged lepton, missing transverse energy, and jets, have been used by several CDF analyses to extend their statistical reach. In Section~\ref{section:APrioriProbability} we introduce the general formula, needed in order to combine triggers, for an \emph{a priori} probability that a trigger selects an event. Since there are several known methods for this task, in Section~\ref{section:StandardMethods} we briefly review the standard trigger combination methods: the inclusion, division, and exclusion methods. These combination methods present difficulties when estimates of the resulting systematic uncertainties are required.  With this motivation, in Section~\ref{section:NovelMethod} we introduce a novel trigger combination method where, on an event-by-event (\emph{in situ}) basis, the trigger with the largest \emph{a priori} probability of selecting the event is considered, while all the other available triggers are ignored. The method was developed to combine these three triggers at CDF in the search for the Standard Model Higgs boson. However, the technique is generic to combinations of any trigger type and any particle physics experiment. In Section~\ref{section:Comparison} we compare the inclusion and novel \emph{in situ} methods in the context of signal event yields in the $WH$ search at CDF. We offer our conclusions in Section~\ref{section:Conclusions}.

\section{Basic Definitions}
\label{section:Definitions}

\ \\Typical collider physics experiments denote as an \emph{event} the activity recorded in a particle physics detector during a particle bunch crossing. Data event samples collected in periods of stable collider and detector conditions are called \emph{runs}. Due to the limited rate of storing events, collider detectors typically use multi-level trigger systems to select online the most interesting and useful data events.

\ \\In this paper we denote L$i$ the $i^{\rm th}$ element from a set of $N$ trigger levels. In a particle physics experiment N is typically 3, although there are exceptions. We denote the smallest unit of a trigger selection as a \emph{trigger object}, for example a requirement on the missing transverse energy or on the transverse energy of a jet. At a given trigger level a \emph{trigger item} is made of a logical {\sc and} between several trigger objects. We define a \emph{trigger chain}, or a \emph{trigger path}, or simply a \emph{trigger}, as the {\sc and} combination of the requirements at each trigger level for a specific event selection. If an event is received as input by one trigger level of one trigger chain and meets the selection criteria of the trigger item of that trigger level, the event is said to have \emph{fired} the trigger level and is typically sent as an input to the next higher trigger level. Given that only a limited number of events can be saved each second, the rates of some lower priority trigger chains are artificially reduced. We define the \emph{prescale} factor $p$ such that, on average, only the $p^{\rm th}$ event that fires a trigger level is sent to the next trigger level. If an event fires a given trigger level and gets sent to the next trigger level, the event is said to have been \emph{taken} by the former trigger level. The prescale of each trigger level can be modified manually between runs and kept constant for all events in a given run; alternatively, it can change automatically for events within a given run, decreasing with the collider's diminishing instantaneous luminosity (\emph{dynamic prescaling}). Prescale decisions can be deterministic or non-deterministic. Since at CDF all prescales are non-deterministic, we consider only this latter category in this paper. Typically all the highest and most of the lower trigger levels are unprescaled, \emph{i.e.} their prescale is unity. If an event is taken at each trigger level of a trigger chain, the event is said to be taken by that trigger and stored for offline analysis. 

\ \\In principle, not all trigger chains designed by an experiment are available for a given event. A trigger chain is \emph{defined} for an event if each of its trigger level requirements is available for that event. Obviously, if a trigger chain is not defined, there is zero chance that it can accept the event. The list of available trigger chains may usually be altered only between runs, so events in a given run have access to the same triggers. For a given event, we denote as $d$ the factor that is unity if the trigger chain is defined for the run, and zero otherwise.

\ \\Each trigger chain is efficient only in a certain region of the kinematic phase space. When analyzing data offline, in order to permit a requirement on the firing of a certain trigger chain, events have to pass a certain trigger-chain-specific event selection. For a given event, we therefore denote as $s$ the factor that is unity if the event passes the offline selection cuts required for the usage of that trigger chain, and zero otherwise.

\ \\For a given offline selection, on an event-by-event basis and for a given trigger, we denote as $e$ the efficiency of one of the trigger levels, \emph{i.e.} the conditional probability that the trigger level fires given that the event is received as input at the same trigger level. For the first trigger level, the event is received as an input if the trigger is defined. For the other trigger levels, the event is received as an input if the event is taken at the previous trigger level. By definition, the efficiencies depend on the offline selection, since they depend on the event kinematics. The efficiency of a trigger path is the product of the efficiencies of the $N$ trigger levels.

\ \\For a given event, we denote as $P$ the \emph{a priori} probability that the event is taken by the entire trigger chain. As will be detailed later in the paper, $P$ takes into consideration not only the trigger level efficiencies and prescales, and whether the trigger chain is defined or not, but also the area of the kinematic phase space that is covered by the trigger chain. In this paper we focus on the case where more than one trigger chain is combined. The total event \emph{a priori} probability to be taken by the desired trigger chain combination is a function of the chosen method and the individual trigger chain \emph{a priori} probabilities. 

\section{Three MET-based Triggers at CDF}
\label{section:Triggers}

\ \\The premise for introducing a novel trigger combination method (Section~\ref{section:NovelMethod}) was that none of the three standard trigger combination methods (Section~\ref{section:StandardMethods}) was optimal at combining the three missing-energy-and-jets trigger chains needed to enhance the statistical sensitivity of the $WH$ search at CDF. We emphasize that the new method presented here has a general applicability to other triggers and experiments.  In order to illustrate the limitations of the standard methods and advantages of the novel technique, we will first introduce the $WH$ analysis and describe some relevant characteristics of the three trigger chains employed.

\ \\The CDF detector is described in detail elsewhere~\cite{CDF}. Given the apparatus geometry, a cylindrical coordinate system is employed, where the $z$ axis points in the proton beam direction, the azimuthal angle is denoted $\phi$, and the polar angle, denoted $\theta$, is expressed in terms of the pseudorapidity $\eta$, defined as $\eta=-\ln(\tan(\theta/2))$. Since the energy (momentum) of a particle is denoted as $E$ ($p$), the transverse component is defined as $\et=E\sin\theta$ ($\pt=p\sin\theta$). A distance in the $\eta-\phi$ plane between two particles is denoted $\Delta R$, defined as $\Delta R=\sqrt{(\Delta \eta)^2+(\Delta \phi)^2}$. The missing transverse energy (MET, or $\met$) is reconstructed and computed as the absolute value of the opposite of the vector sum of all the calorimeter tower energy deposits projected on the transverse plane. Its physical interpretation is the transverse momentum of particles, such as neutrinos, that escape the CDF apparatus undetected. 

\subsection{$WH$ Associated Higgs Search}
\label{subsection:WH}

\ \\We consider the CDF search for the associated production of a Standard Model Higgs ($H$) boson and $W$ boson, where the Higgs boson decays to a bottom quark-antiquark pair and the $W$ boson decays to a charged lepton and a neutrino ($WH \rightarrow l\nu b\bar{b}$). Since a quark is reconstructed in the detector as a shower of collimated particles called a jet, and the neutrino escapes detection, the signature of such a process is a charged lepton, $\met$, and two jets originating from bottom ($b$) quarks. In the offline selection, the charged leptons were reconstructed using stringent criteria (constituting tight charged leptons). Such events are selected using the highly efficient charged-lepton-inclusive triggers (electron and muon). 

\ \\The statistical sensitivity of the analysis can be improved with the addition of an orthogonal sample of events containing charged leptons reconstructed using relaxed criteria (loose charged leptons). The new sample, consisting about 85\% from muon candidates, may be selected at CDF using three trigger chains based on orthogonal kinematic information in the event, namely the MET-and-jets information. We denote these trigger chains, or simply triggers, as MET2J, MET45, and METDI. In this respect, we note that the $\met$ quantity has different values when reconstructed at the L1, L2, or L3 trigger levels, or offline. Combining the three MET-based trigger chains would increase the dataset more than just using one trigger chain alone. In the following subsections we describe the three trigger chains, and their parametrizations, using the full dataset of $9.45 \pm 0.57\ \invfb$ of integrated luminosity recorded by the CDF detector. 		

\subsection{Trigger Descriptions}
\label{subsection:description}

\ \\The MET + 2 jets (MET2J) trigger chain was active for 99.9\% of the integrated luminosity of the full CDF dataset. A data event fires its L1 trigger level if it has $\met > 28\ \gev$ and is then automatically taken at L1 (as there is no prescale at L1). The event is then studied at L2, where it fires if it has $\met > 30\ \gev$ and at least two reconstructed jets, one with a transverse energy $\et > 20\ \gev$ and located in the central region of the detector ($|\eta|<1.1$), while the other jet has $\et > 15\ \gev$ and $|\eta|<2.0$. This trigger chain is prescaled at L2, which means that not all the events that pass requirements at L2 (are fired at L2) are sent to be analyzed at L3 (are taken at L2). The prescale is applied automatically as a function of the instantaneous luminosity. Events that reach L3 and meet the requirement of $\met > 35\ \gev$ fire the trigger chain and, since there is no prescale at L3, also are taken at L3. Since L3 is the last trigger level, the event is taken by the full trigger chain and is saved on tape for offline analysis. The above attributes constitute the last of four major versions of the MET2J trigger chain, which evolved as the instantaneous luminosity increased and the physics analysis requirements changed.  

\ \\The MET-only (MET45) trigger chain had requirements only on $\met$, and not on jets. 
A first version was used for the initial 2.3~$\invfb$ of integrated luminosity and required $\met > 25\ \gev$ at L1, $\met > 25\ \gev$ at L2, and $\met > 45\ \gev$ at L3. A second version increased the cut value at L1 to $\met > 28\ \gev$ and at L2 to $\met > 35\ \gev$, but decreased it at L3 to $\met > 40\ \gev$ in order to select more events from rare processes, such as Higgs production or physics beyond the Standard Model. This trigger chain was never prescaled at any trigger level. 
In the early data taking, 
which corresponds to about $0.16 \pm 0.01\ \invfb$, there was a bug in this trigger chain. The trigger information could not be trusted and data analyses were required to ignore events for this running period. For this reason, in this paper we consider a trigger chain to be undefined if it was operating with known bugs. 

\ \\The third and last MET-based trigger at CDF is the MET + dijet trigger chain. As its name suggests, it is very similar to the MET2J trigger chain, so is denoted as METDI to avoid confusion. The trigger chain was specifically optimized to increase the rate of Higgs boson candidate events. The trigger chain was first introduced when about $2.4 \pm 0.1\ \invfb$ of integrated luminosity had already been collected; it was never modified and never prescaled at any trigger level.

\ \\Since not all trigger chains were defined at the same time, we measured the fraction of integrated luminosity for which each combination of the three MET-based trigger chains was defined, as shown in Table~\ref{table:IntegratedLuminositiesMETTriggers} for the total integrated luminosity of 9.45~$\invfb$.  

\begin{table}[h]
\begin{center}
\begin{tabular}{ccccc}
\hline
MET2J & MET45 &  METDI & Fraction & Integrated \\
\hline
0 & 0 & 0 & 0.000 & 0.000\\
0 & 0 & 1 & 0.000 & 0.000\\
0 & 1 & 0 & 0.001 & 0.001\\
0 & 1 & 1 & 0.000 & 0.001\\
1 & 0 & 0 & 0.017 & 0.018\\
1 & 0 & 1 & 0.000 & 0.018\\
1 & 1 & 0 & 0.233 & 0.251\\
1 & 1 & 1 & 0.749 & 1.000\\
\hline
\end{tabular}
\caption{Fractions and integrated fractions of the CDF full dataset integrated luminosity of 9.45 $\invfb$ for each of the eight combinations given by the possibility that one trigger chain is defined (1) or not defined (0). The fraction of 0.017 of the total integrated luminosity corresponds to the runs where the MET45 trigger chain had a bug and was therefore ignored and considered undefined.}
\label{table:IntegratedLuminositiesMETTriggers}
\end{center}
\end{table}

\ \\Since the MET45 and METDI trigger chains are unprescaled, their total prescales are $1.00 \pm 0.00$. While the MET2J trigger chain is prescaled only at the L2 trigger level, its total prescale equals its L2 prescale. Due to the fact that the L2 prescale was applied only for a fraction of the integrated luminosity and, furthermore, for part of it dynamically as a function of the instantaneous luminosity, we compute an average L2 prescale for the MET2J trigger chain. We use as a reference a trigger chain that was defined for all runs, had no known bugs, and was never prescaled, namely the trigger chain that required at L3 a jet with $\et > 100\ \gev$ (JET100). We computed the average MET2J total prescale as the ratio of the total integrated luminosities of the MET2J and JET100 trigger chains. The average MET2J prescale for the total integrated luminosity of 9.45 $\invfb$, corresponding to the full Tevatron Run II dataset and used in the CDF $WH$ search, has a value of 
$1.10 \pm 0.01$.
The same method confirmed that the MET45 and METDI trigger chains were indeed never prescaled. We also computed the average MET2J prescales for the four different MET2J versions, spanning the full CDF dataset. For the last version, we computed the prescale for the data ranges before and after the METDI trigger chain was put into service. These values are presented in Table~\ref{table:PrescalesMETTriggers}. 

\begin{table}[h]
\begin{center}
\begin{tabular}{cccccc}
\hline
Version & Run Range & Lumi ($\invfb$) & Fraction & Integrated & Prescale\\
\hline
V1  & $<$ 195449           & 0.47 & 0.050 & 0.050 & 1.000\\
V2  &195449-222885            & 0.78 & 0.083 & 0.133 & 1.000\\
V3  &222885-235389           & 0.58 & 0.061 & 0.194 & 1.062\\ 
V4 before METDI & 235389-250000 & 0.47 & 0.050 & 0.244 & 1.091\\ 
V4 during METDI & $>$ 250000 & 7.15 & 0.757 & 1.000 & 1.126\\ 
\hline
\end{tabular}
\caption{Integrated luminosity and average prescales for the MET2J trigger chain. Each row refers to subsequent versions of the trigger chain. The last version is presented before and after the METDI trigger chain was introduced. The columns present, for the run ranges where that trigger version was active, the total integrated luminosity, the fraction of the final dataset integrated luminosity, the integrated fraction of the final dataset luminosity, and the average prescale.}
\label{table:PrescalesMETTriggers}
\end{center}
\end{table}

\ \\The trigger information presented in this subsection is needed in order to model the trigger chains in Monte Carlo simulated events, but remains independent of any analysis that could use these triggers. Other information needed to model the trigger chains, the trigger inefficiencies at each trigger level, depends on the chosen offline event selection of the analysis. In the following, we exemplify the trigger efficiency parametrization measurement using the offline event selection used in several CDF analyses that have the same signature of a charged lepton, missing transverse energy, and two (or more) jets: $WH$, $t\bar{t}H$, $WZ$, and technicolor searches. 

\begin{figure}[htbp]
  \begin{center}
    \includegraphics[width=12.0cm]{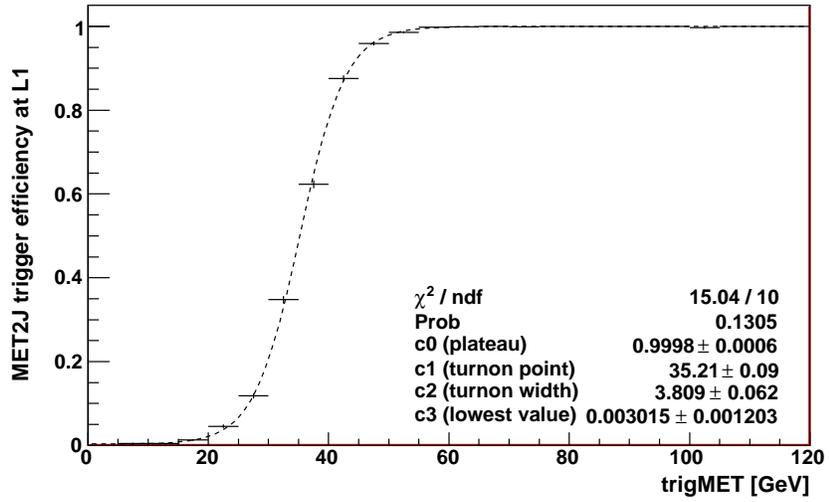}
    \includegraphics[width=12.0cm]{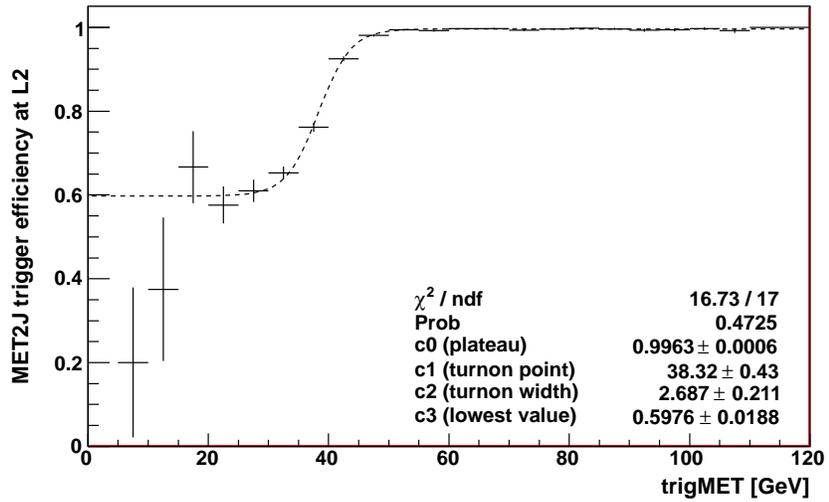}
    \includegraphics[width=12.0cm]{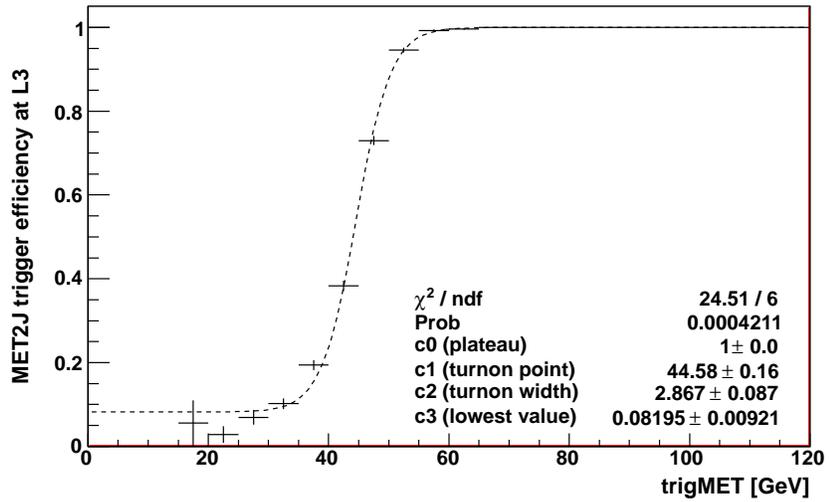}
    \caption{MET2J trigger chain efficiency turn-on curves, parametrized as a function of trigMET, using 9.45 $\invfb$ of integrated luminosity. The figures represent, from top to bottom, the L1, L2, and L3 trigger levels, respectively. The turn-on curves do not include the effect of the prescale for the MET2J trigger chain.}
    \label{figure:ParametrizationMET2J}
  \end{center}
\end{figure}

\subsection{Trigger Parametrizations}
\label{subsection:paramaterization}

\ \\Given the event signature chosen for study, we consider the offline event selection with one muon candidate with $\pt > 20\ \gev$ and $|\eta|<1.1$, and two jets with $\et > 20\ \gev$ and $|\eta|<2.0$. We select a sample of such events using the muon inclusive trigger. The average efficiency for a trigger level belonging to one trigger chain may be measured as the fraction of events taken at the previous trigger level (or defined for the first trigger level) that fire also at the desired trigger level. However, since we want an efficiency evaluation on an event-by-event basis, an average efficiency is not sufficiently precise. Since all three triggers have requirements based on $\met$, we choose to apply the above technique to parametrize the trigger efficiency turn-on curve as a function of the offline $\met$, while applying offline cuts with respect to the jet quantities based on the kinematic distributions of the two jets in the event and employed by the triggers, in order to be in the efficiency plateau region with respect to these variables. 


\ \\We identify for each trigger chain the specific jet cuts that render, for the remaining data events, the turn-on curve parametrization to be flat in any jet quantities and dependent only on trigMET, defined as the raw offline $\met$ corrected for the position of the primary interaction vertex and for the jet energies in the event. All such cuts must be the same as or more stringent than the basic selection of two jets with $\et > 20\ \gev$ and $|\eta|<2.0$. Since the MET45 trigger chain does not have any jet requirements, its offline selection is the basic selection. For the MET2J trigger chain, we have evaluated the turn-on curves for several jet selections and identified the minimal cuts. We require the events to have two jets with $\et > 25\ \gev$, one jet with $|\eta|<0.9$ in order to be in the central region of the detector, and the second jet with $|\eta|<2.0$. The two jets also must have $\Delta R > 1.0$ in order not to overlap in the $\eta$-$\phi$ plane. For the METDI trigger chain, we require events to have two jets with $|\eta|<2.0$, the most energetic jet to have $\et > 40\ \gev$, and the second most energetic jet to have $\et > 25\ \gev$. We can therefore consider that each trigger chain is defined only if the event passes the trigger-specific jet selection.


\ \\Since there are three MET-based trigger chains, each with three trigger levels, for a given integrated luminosity we measure nine trigger efficiency turn-on curves. We demonstrate here the procedure for a generic case. For a given trigger, we select the subset of events from the muon-trigger sample that pass basic event selection and the trigger-chain-specific jet selection. We apply a further selection depending on the trigger level. At L1, we ask that the trigger chain to which it belongs be defined. At L2, we ask that the event was taken at L1 by the correct trigger for this trigger chain. This ensures that the measured efficiency correctly accounts for the fact
that no volunteers are allowed between chains and separates our potential prescales,
computed separately. Similarly, at L3 we ask that the event was taken at L2 by the correct trigger for this trigger chain. For the remaining events we fill the denominator histogram with the variable trigMET. We fill the numerator histogram with the same variable trigMET, only for the fraction of events that also fire the desired trigger chain at the desired trigger level. Since the muon-based trigger chain used to collect the sample of events and the MET-based trigger chain we want to parametrize are uncorrelated for all practical purposes, we divide the numerator and denominator histograms to obtain the efficiency turn-on histogram for the chosen trigger chain at the desired trigger level. We then fit the efficiency histogram to a sigmoid function with four parameters as a function of trigMET, given by

\begin{equation} 
e\rm{(trigMET)}=c_3+\frac{c_0-c_3}{1+e^{-\frac{\rm{trigMET}-c_1}{c_2}}}\ \rm{,}
\label{formula:SigmoidFunction}
\end{equation}

\ \\where $c_0$ represents the highest plateau efficiency, $c_1$ represents the central value of the turn-on region (and is measured in $\gev$), $c_2$ represents the width of the turn-on region (and is also measured in $\gev$), and $c_3$ represents the lowest efficiency value. The fit returns the four parameters that uniquely define the efficiency as a function of trigMET only for a given trigger chain and trigger level. The fit parameter values for the nine turn-on curves obtained using data corresponding to the full CDF dataset are summarized in Table~\ref{table:TurnonCurvesFitParameters}. As an example, Figure~\ref{figure:ParametrizationMET2J} (MET2J) depicts the turn-on curves leading to the first three (MET2J) rows of Table~\ref{table:TurnonCurvesFitParameters}.
We note that all turn-on curves start from values close to zero, as expected, except for the L2 of the MET2J trigger chain, which starts from a value close to 0.6. This is explained by the fact that we average over the full dataset. For the first part of the integrated luminosity there was no L2 requirement for this trigger chain, leading to an efficiency of 1.0. In the latter part of the integrated luminosity, a L2 requirement gave a regular efficiency starting near 0.0.  

\begin{table}[t]
\centering
\begin{tabular}{ccllll}
\hline
Trigger chain & Trigger level & $c_0$ & $c_1$~[GeV] & $c_2$~[GeV] & $c_3$\\
\hline
MET2J & L1 & 1.0 & 35 & 3.8 & 0.0 \\
MET2J & L2 & 1.0 & 38 & 2.7 & 0.6 \\
MET2J & L3 & 1.0 & 45 & 2.9 & 0.1 \\
\hline
MET45 & L1 & 1.0 & 37 & 4.0 & 0.0 \\
MET45 & L2 & 1.0 & 51 & 3.9 & 0.1 \\
MET45 & L3 & 1.0 & 51 & 4.0 & 0.1 \\
\hline
METDI & L1 & 1.0 & 32 & 4.8 & 0.0 \\
METDI & L2 & 1.0 & 29 & 3.3 & 0.1 \\
METDI & L3 & 1.0 & 37 & 2.9 & 0.3 \\
\hline
\end{tabular}
\caption{The fit parameters for the measured turn-on curves, for each trigger level and for each MET-based trigger chain, using 9.45 $\invfb$ of integrated luminosity.}
\label{table:TurnonCurvesFitParameters}
\end{table} 

%

\subsection{Trigger Parametrization Systematic Uncertainties}
\label{subsection:paramaterizationSystematics}

\ \\We note that, unlike the prescale value, which is measured, the efficiency is modelled with a function. There are three types of systematic uncertainties associated with this process. 

\ \\The first uncertainty arises if the shape of the efficiency turn-on curve depends on other variables besides the one we are currently using. In order to study this effect, we select a number of variables on which the efficiency might depend:  $\et$, $\eta$ and $\phi$ of both jets, absolute value of the $\Delta \eta$, $\Delta \phi$, $\Delta R$, $\Delta \et$, as well as fully corrected $\met$ of the analysis, and the fraction of total luminosity that corresponds to each run. We then evaluate the efficiencies in bins of each variable. We repeated the analysis for all these different choices and used the variations we obtained as a measure of the systematic error. The number of bins was chosen automatically to maintain a minimum specified number of events in the turn-on region of the distribution so that the fit could be performed correctly. While for a given event the trigger efficiency was obtained using the corresponding central turn-on curves from Table~\ref{table:TurnonCurvesFitParameters}, the efficiency corresponding to a certain systematic variation was computed using the turn-on curve of the bin in that variable to which the current event belongs. We evaluate the systematic uncertainty on all turn-on curves by picking the largest percentage deviation of the sum of efficiencies, considering all systematic variations from the sum of efficiencies of the central turn-on curves. We obtain a 3\% systematic uncertainty, which we quote on the total yield of Monte Carlo processes (signal and background) in analyses that use these MET-based trigger chains.

\ \\The second type of systematic uncertainty that may affect the trigger efficiency calculation originates in the possibility that the chosen function does not approximate correctly the shape of the efficiency turn-on curve. We studied this effect by varying the fit values of turn-on curves within their statistical errors. The impact on the final result was negligible, being one order of magnitude smaller than the 3\% effect of the first type of systematic uncertainty. In addition, one could consider fitting functions more complex than the sigmoid function, as for example the data is not fully modelled by the sigmoid function at large trigMET. For this reason, we also tested a multivariate regression technique that is able to fit any smooth function. The result was a marginally improved $WH$ background modelling, but a marginally decreased search sensitivity, which lead the CDF collaboration to prefer the better established sigmoid function. Further studies following multivariate fit techniques may be worth pursuing at ongoing particle physics experiments. The fit quality may in principle be improved with a larger dataset to offer higher statistics. However, our dataset could not be extended further as the CDF detector and the Tevatron accelerator were shut down on 30 September 2011. In this paper we employ the full CDF dataset. 

\ \\The third type of efficiency-affecting systematic uncertainty is the mismodelling in simulation of the quantities used in the trigger efficiency parameterization. We check this in the broader analysis. Since the quantities used are well modelled, this effect is negligible.

\section{Trigger Chain \emph{in situ} \emph{a priori} Probability}
\label{section:APrioriProbability}

\ \\As discussed in Section~\ref{section:Triggers}, on an event-by-event basis a trigger chain may be defined or not, may have different efficiencies and prescales at each available trigger level, and may be applied or not depending on whether the event falls in the kinematic phase space applicable to the trigger chain. For each of the trigger chains available, we compute, \emph{in situ}, an \emph{a priori} probability that the trigger chain is taken for the event. We introduce a formula that has the flexibility to model the complexity of the trigger chains described in Section~\ref{section:Triggers}:

\begin{equation} 
P_{\rm trigger}=d \cdot s \cdot \prod_{i=1}^{i=N_{\rm trigger\ level}}(e_i \cdot \frac{1}{p_i})\rm{,}
\label{formula:ProbabilityTriggerInSitu}
\end{equation}

\ \\where $d$ is unity if the trigger chain is defined for the run to which the event belongs and zero if it is not; $s$ is unity if the event passes the event selection specific to the trigger chain and zero if it does not; $e_i$ and $p_i$ are the efficiency and prescale for the $i^{\rm th}$ trigger level of the trigger chain, and $N_{\rm trigger\ level}$ is the total number of trigger levels. The minimum value of $P_{\rm trigger}$ is zero and the maximum value is unity. We exemplify how $P_{\rm trigger}$ is computed \emph{in situ} in the case of the three MET-based trigger chains, for both data and MC-simulated events, in the full CDF dataset spanning 9.45~$\invfb$ of integrated luminosity.

\ \\For the value of $d$, one needs to know if the trigger chain is defined for the run to which the event belongs. In case of a bug, as existed for the MET45 trigger chain, the trigger chain should be considered undefined even for a particular run range where the trigger chain is defined, as described in Subsection~\ref{subsection:description}. For each data event, it is known if the trigger is defined or not based on the run number. 
For MC simulated events at CDF there is no trigger simulation, so run numbers do not correspond to real data run numbers and therefore should be ignored. As a result we simulate the fraction of the total integrated luminosity to which an event belongs by throwing a random number from a set uniformly distributed between 0 and 1. By comparing this number with the numbers in the last column of Table~\ref{table:IntegratedLuminositiesMETTriggers}, we simulate which trigger chains are considered to be defined for that event. 

\ \\It is the value of $s$ that permits the consideration of trigger chains that span different regions of kinematic phase space. If the event falls in the kinematic region where the trigger chain is in the plateau of the efficiency with respect to quantities other than that used to parametrize the triggers, then the value of $s$ is one; otherwise it is zero. In the case of the MET-based trigger chains, the trigger-chain-specific event selection involves jet quantities only and the specific cuts are described in Subsection~\ref{subsection:paramaterization}. This value is computed the same way for data and MC-simulated events.

\ \\We note that the formula in Eq.~\ref{formula:ProbabilityTriggerInSitu} scales with finite number of trigger levels. For each trigger level we compute an efficiency value $e$ and a prescale value $p$. The efficiency value $e$ is computed using the sigmoid function of trigMET described by Eq.~\ref{formula:SigmoidFunction}. The four parameters of the function are specific to each trigger chain and trigger level and are listed in Table~\ref{table:TurnonCurvesFitParameters}. All of the MET-based trigger chains are unprescaled at all trigger levels (so we consider the value of $p$ to be unity), except in the case of the L2 of the MET2J trigger chain. For each event we use Table~\ref{table:PrescalesMETTriggers} to find out to which version of the MET2J trigger chain the event belongs and we apply the appropriate prescale value from the last column of the table. For data events we use the run number and for MC simulated events we use the already computed random number simulating the fraction of the total integrated luminosity to which the event belongs. 

\ \\Using Eq.~\ref{formula:ProbabilityTriggerInSitu} we compute for each event, either data or MC-simulated, three \emph{a priori} probabilities that each of the three MET-based trigger chains selects the event. We are now ready to combine the trigger chains using different methods. We will first review the standard methods and then introduce the novel \emph{in situ} combination method.  

\section{Review of Existing Trigger Combination Methods}
\label{section:StandardMethods}

\ \\Several standard methods to combine data samples selected with independent trigger requirements are presented in detail by a paper inspired by work from the H1 collaboration at the HERA accelerator~\cite{ReviewTriggerCombinationNIM}: the inclusion, division, and exclusion methods. In this section we briefly review the key points of each method. We note that in Ref.~\cite{ReviewTriggerCombinationNIM} the authors weight data events in order to match with MC simulated events, whereas we take the approach of weighting MC events in order to match with data events. 

\subsection{Inclusion Method}
\label{subsection:inclusion}

\ \\The inclusion method is an ideal, but unrealistic, event-by-event (\emph{in situ}) trigger chain combination method. It performs a logical '{\sc or}' between all available trigger chains. On an event-by-event basis, a data event is checked if it is taken by at least one of the available trigger chains. If it is, the event is accepted (it is given a weight of 1.0); if it is not, the event is rejected (it is given a weight of 0.0). In contrast, a MC-simulated event is, on an event-by-event basis, always accepted and assigned a fractional weight (between 0.0 and 1.0) equal to the \emph{a priori} probability $P_{\rm{inclusion}}$ that the event is taken by the {\sc or} between the available trigger chains. Since the only case where the {\sc or} method fails is when none of the trigger chains selects the event, if there is no correlation between trigger chains, $P_{\rm{inclusion}}$ is given by the formula

\begin{equation} 
P_{\rm{inclusion}}=1-\prod_{i=1}^{i=N_{\rm trig}} (1-P_{i}) \rm{,}
\label{formula:APrioriPTotalInclusion}
\end{equation}

\ \\where $P_i$ is the \emph{a priori} probability that the $i^{\rm th}$ trigger chain is taken for the event, given by Eq.~\ref{formula:ProbabilityTriggerInSitu}, and $N_{\rm trig}$ is the number of available trigger chains. If there are correlations, as in the case of the three MET-based trigger chains at CDF, evaluating a systematic uncertainty is not sufficient and Eq.~\ref{formula:APrioriPTotalInclusion} would have to be extended further. These two points make the inclusion method impractical. 

\ \\We note that the description of the inclusion method in Ref.~\cite{ReviewTriggerCombinationNIM} may be summarized as an \emph{a priori} probability given by a formula identical to that in Eq.~\ref{formula:ProbabilityTriggerInSitu}, only assuming that $s=1$, as shown in the formula

\begin{equation} 
P_{\rm trigger\ inclusion}=d \cdot \prod_{i=1}^{i=N_{\rm trigger\ level}}(e_i \cdot \frac{1}{p_i})\rm{.}
\label{formula:ProbabilityTriggerInclusionExclusion}
\end{equation}

\ \\We introduced the term $s$ as it is crucial when combining trigger chains that span different regions of phase space, thus further increasing the statistical power. Besides the slightly different regions in the MET and jets phase space evident in our example, one can also imagine combining electron and muon trigger chains, where an electron (muon) candidate event would have $s=1$ for the electron (muon) trigger and $s=0$ for the muon (electron) trigger. In this way one could combine the electron and muon samples into a single sample, thus reducing the analysis bookkeeping, while also ensuring the best statistics available. 

\ \\The major advantage of this method is that it is scalable to any number of triggers.  A second advantage is that, by definition, the inclusion method offers the largest event yield possible, both for data and MC-simulated events. A third advantage is that it is an \emph{in situ} method, which means that no studies have to be performed prior to the physical analysis. For these reasons the inclusion method is an ideal technique. 

\ \\The method, however, also presents a major disadvantage. Since for data events it checks for each trigger chain if it takes the event or not in order to compute the logical {\sc or}, correlations between the trigger efficiencies must be taken into account. Computing the systematic uncertainty for the correlation is not scalable with the number of trigger chains, though, as the calculation becomes increasingly complex as more trigger chains are considered. Furthermore, the sensitivity of the physics analysis decreases when a new systematic uncertainty is introduced. 

\subsection{Division Method}
\label{subsection:Division}

\ \\Another standard method is the division method. Here the phase space is divided into several orthogonal kinematic regions such that only the trigger chain that is, on average, optimum in a given kinematic region is assigned to all events in that region. Since all other trigger chains are totally ignored once the trigger chain choice has been made, this method lacks the disadvantage of the inclusion method, as all correlations between trigger chains are avoided. If this were not the case, one would have to introduce a new systematic uncertainty, which would be both difficult to compute and potentially large in value in the case of nearly fully correlated trigger chains. The division into orthogonal kinematic regions is done by recursive studies performed offline, but before the main analysis. As a result, the division method is not an \emph{in situ} method. The study must be repeated every time the offline event selection is changed, which is not the case for an \emph{in situ} method. The kinematic regions are determined by trial and error so that the average \emph{a priori} probability in one region for one trigger chain dominates that of other trigger chains. 

\ \\In this context, the formula for one trigger chain \emph{a priori} probability can only take into account the efficiencies at all trigger levels, while assuming the trigger chains are not prescaled ($p_i=1$), and that all trigger chains are defined ($d=1$). Furthermore, for more than two trigger chains, the entire procedure becomes very complicated, so the method does not scale with any finite number of trigger chains, like an \emph{in situ} method does. The text in Ref.~\cite{ReviewTriggerCombinationNIM} can be summarized in the following formula for the \emph{a priori} probability:

\begin{equation} 
P_{\rm trigger\ division}=\prod_{i=1}^{i=N_{\rm trigger\ level}}(e_i)\rm{,}
\label{formula:ProbabilityTriggerDivision}
\end{equation}

\ \\which clearly shows the method's limiting assumptions. 

\ \\During the main analysis, on an event-by-event basis, a data event is checked if it is taken by the chosen trigger. If it is, the event is accepted (it is given a weight of 1.0); if it is not, the event is rejected (it is given a weight of 0.0), without checking the other triggers. A MC-simulated event, however, is always accepted and assigned a fractional weight (between 0.0 and 1.0) equal to the \emph{a priori} probability $P_{\rm trigger\ division}$ that the event is taken by the chosen trigger.


\subsection{Exclusion Method}
\label{subsection:Exclusion}

\ \\The third and last standard method described in Ref.~\cite{ReviewTriggerCombinationNIM} is the exclusion method. While the division method divides the event phase space into orthogonal regions based on the kinematic information of the events, the exclusion method does so based on information about whether the trigger chains have fired the events. It is an \emph{in situ} method. On an event-by-event basis, all available triggers are checked if they have fired the event. If they have not, they are ignored. From the trigger chains that have, the trigger chain with the largest \emph{a priori} probability given by Eq.~\ref{formula:ProbabilityTriggerInclusionExclusion} is assigned to the event. We note that this formula includes terms specifying whether the trigger chain is defined, its prescales, and its efficiencies, as this information is needed to determine if the trigger chain has fired or not.

\ \\Once a trigger chain is assigned to each event, the same procedure as for the division method applies. In the ideal case one would have a working and exact trigger simulation. However, it is in general difficult to have a working and reliable trigger simulation, particularly in the high-occupancy environment of a hadron collider and when using triggers involving the hadronic final state. Therefore, since it is checked whether the trigger chains have fired before a single trigger chain is assigned to each event, one needs to introduce a systematic uncertainty due to the correlations between triggers.  As for the inclusion method, the requirement of understanding this uncertainty renders this technique impractical.


\section{A Novel \emph{in situ} Trigger Combination Method}
\label{section:NovelMethod}

\ \\Motivated by our CDF Higgs search example, and equipped with an understanding of existing techniques, we now introduce a new trigger chain combination method. It generalizes both the division and exclusion methods, as the phase space is both divided into orthogonal kinematic and trigger regions. This is achieved by considering each event as an independent region of phase space. As in both the division and exclusion methods, only one trigger chain is assigned to each event. It is also an \emph{in situ} method. During the main analysis, on an event-by-event basis, an \emph{a priori} probability $P$, given by Eq.~\ref{formula:ProbabilityTriggerInSitu}, that the event is taken is computed for each desired trigger chain, and only the trigger chain with the largest $P$ is assigned to the event. This allows to optimise the combination of efficiency, prescale and the region of definition and utilisation of triggers.

\ \\The method has all the advantages of the \emph{in situ} methods, such as avoiding studies before the main analysis and scaling with any finite number of trigger chains. Furthermore, since the trigger chain is chosen before any trigger chain is checked whether it has fired a data event or not, there are no correlations between trigger chains to take into account, which was a limiting factor for the inclusion and exclusion methods. Moreover, since the chosen trigger chain is the one with the largest \emph{a priori} probability, the \emph{in situ} method has \emph{by definition} the largest event yield possible from all the variants of the division and exclusion methods. It is only the inclusion method that has a higher yield, but a price of correlations between trigger chains. 

\ \\As seen in Table~\ref{table:MethodsComparison}, the novel \emph{in situ} method is the only method that is both an \emph{in situ} method and does not need to take into account correlations between trigger chains. As a result, there is no systematic uncertainty to be taken into account for correlation between triggers. The only systematic uncertainties are on a trigger-by-trigger basis in calculating the trigger efficiency, prescale, and the effect of the random number generator. The method is therefore better suited to combining a large number of trigger chains whose characteristics have changed over several years of data taking, a typical situation for large contemporary particle physics experiments.

\begin{table}[t]
\centering
\begin{tabular}{lll}
\hline
Method & Is \emph{in situ}? & Has correlations between trigger chains?\\
\hline
Inclusion & Yes & Yes\\
Division  & No  & No\\
Exclusion & Yes & Yes\\
Novel     & Yes & No\\
\hline
\end{tabular}
\caption{Comparison between the four trigger combination methods based on whether they are an \emph{in situ} method and if they need to take into account correlations between trigger chains. The inclusion and exclusion methods require full trigger simulation in order to handle trigger correlations. Since the trigger simulation is imperfect, the inclusion and exclusion methods are impractical. On the other hand, the division and novel \emph{in situ} methods do not introduce correlations between trigger chains. }
\label{table:MethodsComparison}
\end{table}

\section{Comparison between the Novel and Inclusion \emph{in situ} Methods}
\label{section:Comparison}

\subsection{Context of the $WH$ Search Example}
\label{subsection:WHSearch}

\ \\Given the characteristics of the three MET-based trigger chains described in Section~\ref{section:Triggers}, it would be very difficult to combine the three trigger chains with any method other than the novel and inclusion \emph{in situ} methods. Furthermore, for the inclusion \emph{in situ} method, it would be difficult to compute the systematic uncertainty arising from trigger chain correlations. We therefore used the novel \emph{in situ} method to combine the three trigger chains in the context of the CDF $WH$ analysis, also described in Section~\ref{section:Triggers}. We can, however, compare the two methods to quantify what percentage of event yield is lost by not using the inclusion method, with the \emph{caveat} that, in order to compute the \emph{in situ} inclusion method yield, we used Eq.~\ref{formula:APrioriPTotalInclusion}, which assumes the trigger chains are uncorrelated. 


\ \\The $WH$ analysis considers a control sample formed by all events that pass the standard event selection, where the signal yield is expected to be comparatively negligible. The signal regions are represented by the subset of the control sample where one or two jets are required to be identified to originate from bottom quarks ($b$ tagging). In this paper we use as the signal region the most sensitive signal region from the $WH$ analysis, where both jets are $b$ tagged using a multivariate $b$ tagger~\cite{HOBIT}. The standard event selection of the WH CDF analysis is provided in Ref.~\cite{WHPRD7.5}~\cite{BuzatuPhDThesis}.

\ \\In this study we consider only a $WH$ Monte Carlo sample produced assuming a Higgs boson mass of $125 \gevcc$.

\begin{figure}[htbp]
  \begin{center}
    \includegraphics[width=12cm]{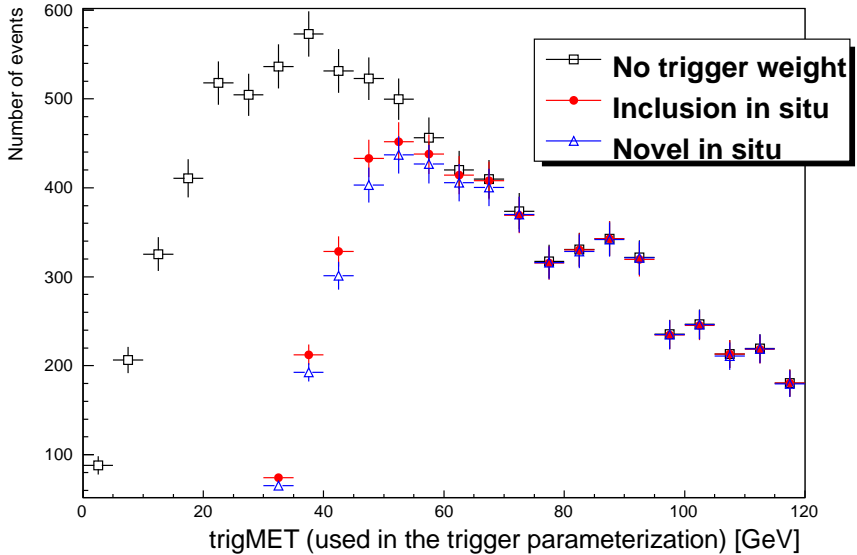}
    \caption{A simulated distribution of the trigMET used to parametrize the trigger efficiencies at each trigger level in the signal region and for isolated charged lepton events. The vertical scale is only proportional to the event yield, as the analysis integrated luminosity and the signal cross-section and branching fraction are not applied. We compare not applying any trigger weight and applying a trigger weight using either the inclusion or novel \emph{in situ} method. We see how the event count decreases when applying the trigger weights and how the inclusion \emph{in situ} method provides a marginally larger event prediction than the novel \emph{in situ} method.}
    \label{figure:Overlaid}
  \end{center}
\end{figure}

\subsection{Trigger Weight Calculation}
\label{subsection:TriggerWeightCalculation}

\ \\In this subsection we outline the calculation of a trigger chain event weight for a MC event and the trigger chain decision for a data event. First we use Eq.~\ref{formula:ProbabilityTriggerInSitu} to compute, on an event-by-event basis and for each of the three MET-based trigger chains, the~\emph{a priori} probability that the event fires for that trigger chain. We use the trigger chain characteristics corresponding to the integrated luminosity of 9.45 $\invfb$ described also in Section~\ref{section:Triggers} .

\ \\In the case of a MC-simulated sample, we simulate the fraction of the total integrated luminosity the event belongs to by selecting a random number from a sample of numbers uniformly distributed between 0 and 1. Depending on this number we decide which trigger chains are defined and what is the MET2J prescale for the event, as described in Tables~\ref{table:IntegratedLuminositiesMETTriggers} and~\ref{table:PrescalesMETTriggers}, thus determining the terms $d$ and $p_i$ from Eq.~\ref{formula:ProbabilityTriggerInSitu}. By changing the random number seed we identify a 0.3\% systematic uncertainty on the event yield, which is negligible compared with the 3\% contribution on the event yield due to the trigger efficiency turn-on curve parametrizations. In the case of a data sample, the trigger chains are considered to be defined or not based on the actual event information. If a trigger chain is known to have a bug for the particular run to which the event belonged, the trigger chain is considered undefined, as was the case for the MET45 trigger chain in the early data taking period.

\ \\For either MC or data samples, we compute the value of the trigger-chain-specific jet selection $s$ used in Eq.~\ref{formula:ProbabilityTriggerInSitu} for the current event by applying the cuts specified in Subsection~\ref{subsection:paramaterization} on the offline-reconstructed variables $\et$ and $\eta$ of each of the two jets, as well as on the $\Delta R$ between the two jets. Depending on the trigger, $s$ may be zero or one for the same event. Furthermore, for each trigger level of each trigger chain, we compute the efficiency term $e_i$ from Eq.~\ref{formula:ProbabilityTriggerInSitu} by substituting in Eq.~\ref{formula:SigmoidFunction} the offline-reconstructed trigMET defined in Subsection~\ref{subsection:paramaterization}. 

\ \\In Figure~\ref{figure:Overlaid} we present a simulated sample of isolated charged lepton events in the signal region. 
As the analysis integrated luminosity and the signal cross-section and branching fraction are not applied, the vertical axis values do not represent the full event yield, but are proportional to it. Therefore, we are able to compare the effect of not applying the trigger weight to that of applying the trigger weight using either the inclusion or the novel \emph{in situ} method. 

\ \\Once we have computed the \emph{a priori} probability $P$ for each of the three MET-based triggers for the current events, we proceed to computing the event weight using the two methods under comparison.

\ \\In the case of the inclusion \emph{in situ} method, for a data event we check if at least one of the three trigger chains has taken the event. If it has, the event is kept (given a weight of 1.0), else the event is rejected (given a weight of 0.0). If the event is simulated using MC, the event is assigned the weight equal to the probability that it is selected by at least one of the three trigger chains, which is computed using Eq.~\ref{formula:APrioriPTotalInclusion}, taking as inputs the $P$ values of each of the three MET-based trigger chains. 

\ \\In the case of the novel \emph{in situ} method, for either data or MC events, we assign to the event only the trigger chain with the largest $P$ value, while ignoring the other trigger chains. If it is a data event, we check if the chosen trigger chain has taken the event. If it has, the event is kept, else it is rejected. If it is a MC event, we consider that the event has fired the assigned trigger chain and we assign to it a weight equal to $P$ of the chosen trigger chain, \emph{i.e.} the largest $P$ for the event. 

\subsection{Yield Comparison}
\label{subsection:YieldComparison}

We are now in a position to compute ratios of event yields of each trigger chain alone and of the trigger combined using the inclusion \emph{in situ} method relative to that combined using the novel \emph{in situ} method. We perform the exercise for both the control sample (pretag) and most sensitive signal sample (double tag). 
The relative event yields are presented in Table~\ref{table:RelativeEventYields}.

\begin{table}[thb]
\centering
\begin{tabular}{lcc}
\hline
Triggers & Control & Signal \\
\hline
MET2J                        & 0.67 & 0.74\\
MET45                        & 0.82 & 0.79\\
METDI                        & 0.65 & 0.70\\
All Inclusion \emph{in situ} & 1.02 & 1.02\\
All Novel \emph{in situ}     & 1.00 & 1.00\\
\hline
\end{tabular}
\caption{Relative event yields in a 125 $\gevcc$ Monte Carlo $WH \rightarrow l\nu b\bar{b}$ simulated sample, where the three MET-based triggers have the characteristics of the full CDF dataset of 9.45~$\invfb$ of integrated luminosity. The rows present different trigger configurations: each trigger used alone, followed by all triggers combined with the inclusion and novel~\emph{in situ}  methods. All ratios are relative to the novel~\emph{in situ} yield.
The 'Control' column refers to the pre-tag sample, whereas the 'Signal' column refers to the 2-$b$-tag sample.}
\label{table:RelativeEventYields}
\end{table} 

\ \\The values from the table illustrate that the MET-based trigger chains are correlated on the order of 60-80\%. Since the trigger weight calculation for the \emph{in situ} inclusion method assumes the trigger chains are not correlated, computing the event yield ratio is the best we could do in order to quantify the \emph{in situ} inclusion method. This emphasizes the impracticality of this method and the versatility of the novel \emph{in situ} method. 

\ \\The novel \emph{in situ} method produces an event yield only 1-2\% lower than the inclusion method (also seen in Figure~\ref{figure:Overlaid}), which is an acceptable cost for eliminating the need of introducing a systematic uncertainty due to the correlation between trigger chains, one that would both decrease the sensitivity of the analysis and would be difficult to compute. We also see that selection using only a single trigger chain produces a yield approximately 20-35\% lower than the combined method. This suggests that the trigger chains are highly correlated with each other, but that they do not overlap fully in the kinematic phase space, which allows the combined method to provide gains significantly beyond one trigger chain alone. We also see that we obtain about the same number in the control (pretag) sample and the signal (2 tag) sample. 

\subsection{Generic Illustrative Case}
\label{subsection:ComparisonGeneric}

\ \\While in the previous subsection we compared the sum of the weights of a sample of events that represented the event yields, we now compare the weights for a single hypothetical generic event for which we assume three triggers are available and defined, each with an \emph{a priori} probability of accepting the event that is either small ($P=0.01$), medium ($P=0.50$), or large ($P=0.99$). In the example presented in Table~\ref{table:GenericComparison}, the trigger path $P_3$ is the best of the three paths entirely by design, since this is a table of combinations and not permutations. The table showcases that, although the $P$ value of the triggers combined with the inclusion \emph{in situ} method changes, the same value for the novel \emph{in situ} method is typically lower by only about 1\%. The statistical limitation of the \emph{in situ} method is for those events where no trigger has a large $P$, as is the case in the first row of Table~\ref{table:GenericComparison}. In typical physics analyses, such events represent only a negligible fraction of the entire event sample. This simple generic example illustrates that, if at least one of the trigger paths has a large $P$, the new method provides almost optimal efficiencies.

\begin{table}[tfb]
\centering
\begin{tabular}{cccccc}
\hline
$P_1$ & $P_2$ & $P_3$ & Inclusion \emph{in situ} & Novel \emph{in situ} & Ratio \\
\hline 
0.01 & 0.01 & 0.01 & 0.030 & 0.010 & 2.97\\
0.01 & 0.01 & 0.50 & 0.510 & 0.500 & 1.02\\
0.01 & 0.01 & 0.99 & 0.990 & 0.990 & 1.00\\
0.01 & 0.50 & 0.50 & 0.753 & 0.500 & 1.51\\
0.01 & 0.50 & 0.99 & 0.995 & 0.990 & 1.01\\
0.01 & 0.99 & 0.99 & 1.000 & 0.990 & 1.01\\
0.50 & 0.50 & 0.50 & 0.875 & 0.500 & 1.75\\
0.50 & 0.50 & 0.99 & 0.998 & 0.990 & 1.01\\
0.50 & 0.99 & 0.99 & 1.000 & 0.990 & 1.01\\
0.99 & 0.99 & 0.99 & 1.000 & 0.990 & 1.01\\
\hline
\end{tabular}
\caption{Comparison between the inclusion and novel \emph{in situ} methods for a hypothetical event with three triggers. Each trigger may have an  \emph{a priori} probability ($P$) of accepting the event that is either small ($P=0.01$), medium ($P=0.50$), or large ($P=0.99$). There are ten such unique combinations. By design, the values $P_1$, $P_2$, $P_3$ are shown in increasing order. We note that the inclusion \emph{in situ} method must assume uncorrelated triggers, which are rare.}
\label{table:GenericComparison}
\end{table}

\subsection{Generic Yield Discussion}
\label{subsection:GenericYield Discussion}

\ \\In general, the yield of the novel \emph{in situ} method is smaller than that of the \emph{in situ} inclusion method, but there is no need to determine trigger correlations. Furthermore, the higher the degree of correlation between triggers, the less yield is lost, and the higher the systematic effect due to correlations would otherwise be.  Also, in regions where individual trigger paths have a substantially higher \emph{a priori} probability, the advantage of the inclusion method is smaller. 

\ \\The main advantage of the novel \emph{in situ} when compared to the division method is that the phase space does not have to be separated manually by iterative studies before the main analysis. If one achieved this separation well, the yield outcome would be the same. Therefore the relative advantage of the new method consists of the time saved by being an \emph{in situ} method. The higher the complexity of the trigger efficiency as a function of phase space, the higher the gain. 

\ \\The exclusion method would be able to accept a larger event yield than the novel \emph{in situ} method, but at the price of needing a perfect trigger simulation or otherwise understanding the correlations between the trigger chains. In practice this is not achievable and the novel method is more practical. 

\ \\In conclusion, the novel \emph{in situ} method works well if there is no trigger simulation or another way to determine the correlations between triggers.  If this were possible, either the \emph{in situ} inclusion or the exclusion method would likely be preferred.

\subsection{Generic Uncertainty Discussion}
\label{subsection:GenericUncertaintyDiscussion}

\ \\If a perfect trigger simulation were available to determine the trigger efficiencies, the first and second type of systematic uncertainty on the trigger efficiency (described in Subsection~\ref{subsection:paramaterizationSystematics}) would no longer be present, as a working trigger simulation usually offers more accuracy than an ad-hoc parametrization function. However, the third type of systematic uncertainty would likely increase, since a trigger simulation depends on trigger level quantities ({\it i.e.} before corrections and calibrations) that are inherently more difficult to simulate and typically don't get checked as rigorously. If the trigger simulation is trusted both for the efficiencies and the trigger correlations, it is best to use the \emph{in situ} inclusion method. For a real experiment such as CDF, there are always systematic uncertainties present and these are correlated between trigger paths. In order to implement the inclusion \emph{in situ} method, these would be difficult to evaluate and introducing them would decrease the sensitivity of a physics search. Using the novel \emph{in situ} method, however, circumvents these problems by avoiding any additional systematic uncertainty, since only one trigger path is selected for use before checking whether the event passed that trigger path or not.


\section{Conclusions}
\label{section:Conclusions}

\ \\In this paper we have reviewed the essential characteristics and uses of triggers for particle physics experiments. As a specific motivating example, we described three MET-based trigger chains available in the full CDF dataset. Their situation is complex as, during the multi-year data taking period, the trigger chains often changed configuration, one trigger chain was introduced later, and another trigger chain had a bug during a part of the data taking. These trigger chains could be considered in the context of the $WH$ search at CDF in order to increase the statistical sensitivity of the analysis, which was subsequently combined with other Tevatron searches to find evidence of a new particle consistent with the Standard Model Higgs boson~\cite{TevatronHiggsEvidence}.

\ \\We reviewed three standard trigger chain combination methods. As an \emph{in situ} method, the inclusion method scales well with any number of trigger chains. Since it considers a logical '{\sc or}' between trigger chains, it produces the highest event yield possible. The method would be ideal were it not impractical due to the need to take into account the correlation between trigger chains by introducing a systematic uncertainty that is both difficult to compute and would decrease the analysis sensitivity. Since the division method is not an \emph{in situ} method, it requires several studies to be performed before the main analysis. It divides the event phase space into orthogonal kinematic regions and assigns only one trigger chain to each event. In this manner, correlations between trigger chains are avoided. The method becomes impractical for three or more trigger chains. The exclusion method is another \emph{in situ} method that divides the event phase space into orthogonal trigger regions. On an event-by-event basis, trigger chains are checked whether they select the event and, from the trigger chains that do, the one with the highest \emph{a priori} probability is assigned to the event. Since trigger chains are checked before one is assigned to the event, this method also needs to take into account the correlation between trigger chains.

\ \\We introduced a novel \emph{in situ} method to combine trigger chains in a manner that generalizes the division and exclusion methods in dividing the event phase space simultaneously into orthogonal kinematic and trigger regions. This is achieved by considering each event as its own independent region, which makes the method an \emph{in situ} one. On an event by event basis, the trigger chain with the largest \emph{a priori} probability is assigned to the event, while the other trigger chains are ignored. This ensures that there is no need to take into account correlations between trigger chains. At the same time, by definition this method provides the highest event yield, with the inclusion method being the only exception. 

\ \\We compared the novel and inclusion \emph{in situ} methods in the context of the CDF $WH$ search, where a new charged lepton category with relaxed criteria was selected with the help of the three MET-based trigger chains. The combined methods provide a yield of about 20-35\% above those for one trigger chain alone. The novel method has a yield of only about 1-2\% lower than the inclusion method, which is an acceptable cost for the benefit of avoiding the correlations between triggers. This small difference is possible because at least one trigger chain was very efficient for the majority of events, as different triggers were sensitive to different kinematic regions of phase space. Combining such trigger chains was made possible by our introduction of a new term in the \emph{a priori} probability in the formulae deduced from the review article in Ref.~\cite{ReviewTriggerCombinationNIM}. The only limitation of the novel \emph{in situ} method is in the rare case where no trigger is highly efficient.  

\ \\Although the method has been introduced in the context of the $WH$ Higgs boson search at CDF, it is generic. 
The described novel {\it in situ} trigger chain combination method will be useful for analyses of different signatures at other experiments, such as those at the Large Hadron Collider, especially after several years of data taking when it will become increasingly important to combine several different run periods under diverse trigger conditions.

\section{Acknowledgements}
\label{section:Acknowledgements .}

\ \\The authors would like to acknowledge the invaluable support provided by the CDF collaboration and the staff from Fermilab and their home institutions. Special thanks are due to P.~Wilson of Fermilab for comments on the manuscript. A.~Buzatu and A.~Warburton were supported by the Natural Sciences and Engineering Research Council of Canada and the Universities Research Association, Inc.~(USA). The contributions of N.~Krumnack and W.~Yao were supported by the U.S. Department of Energy. N.~Krumnack is currently supported by the U.S. ATLAS Operations Program.



\bibliography{reference}
\bibliographystyle{h-elsevier3}

\end{document}